%% file: main.tex
\documentclass[sigconf]{aamas}  % do not change this line!
\pdfoutput=1

\mathchardef\mhy="2D

\usepackage{amsfonts}
\usepackage{amsmath}
\usepackage{tikz}
\usepackage{float}
\usepackage{makecell}
\usepackage{caption}

\usepackage{pgfplots}
\usepackage{pgfplotstable}

\pgfplotsset{compat=1.17}
\begin{document}

\title[A Spatially Dependent House Hunting Model]{A Spatially Dependent Probabilistic Model for House Hunting in Ant Colonies}  

\author{Grace Cai}
\affiliation{%
 \institution{MIT}
 \city{Cambridge} 
 \state{MA} 
}
\email{gracecai@mit.edu}

\author{Wendy Wu}
\affiliation{%
 \institution{MIT}
 \city{Cambridge} 
 \state{MA} 
}
\email{wsw23@mit.edu}

\author{Wayne Zhao}
\affiliation{%
 \institution{MIT}
 \city{Cambridge} 
 \state{MA} 
}
\email{wayzhao@mit.edu}

\author{Jiajia Zhao}
\affiliation{%
 \institution{MIT}
 \city{Cambridge} 
 \state{MA} 
}
\email{jiajiaz@mit.edu}

\author{Nancy Lynch}
\affiliation{%
 \institution{MIT}
 \city{Cambridge} 
 \state{MA} 
}
\email{lynch@csail.mit.edu}

\Large

\begin{abstract}  % put your abstract here!
\textbf{Ant species such as \textit{Temnothorax albipennis} select a new nest site in a distributed fashion that, if modeled correctly, can serve as useful information for site selection algorithms for robotic swarms and other applications. Studying and replicating the ants' house hunting behavior will also illuminate useful distributed strategies that have evolved in nature. Many of the existing models of househunting behaviour for \textit{T. albipennis} make the assumption that all candidate nest sites are equally distant from the ants' home nest, or that an ant has an equal probability of finding each candidate nest site. However, realistically this is not the case, as nests that are further away from the home nest and nests that are difficult to access are less likely to be found, even if they are of higher quality. We extend previous house-hunting models to account for a pairwise distance metric between nests, compare our results to those of real colonies, and use our results to examine the effects of house hunting in nests of different spatial orientations. Our incorporation of distances in the ant model appear to match empircal data in situations where a distance-quality tradeoff between nests is relevant. Furthermore, the model continues to be on par with previous house-hunting models in experiments where all candidate nests are equidistant from the home nest, as is typically assumed.}
\end{abstract}

\maketitle

%%%%%%%%%%%%%%%%%%%%%%%%%%%%%%%%%%%%%%%%%%%%%%%%%%%%%%%%%%%%%%%%%%%%%%%%%%%%%%%%%%%%%%%%%%%%%%%%%%%%%%%%%
%% start of main body of paper

\section{Introduction}\label{sec:introduction}
\input{doc/intro.tex}

\section{Background}\label{sec:background}
\input{doc/background.tex}

\section{Methods}\label{sec:methodology}
\input{doc/methodology.tex}

\section{Experimental Design}\label{sec:setup}
\input{doc/setup.tex}

\section{Results}\label{sec:results}
\input{doc/results.tex}

\section{Discussion}\label{sec:discussion}
\input{doc/discussion.tex}

\section{Future Work}\label{sec:future}
\input{doc/future.tex}

% \begin{acks}

% \end{acks}

%%%%%%%%%%%%%%%%%%%%%%%%%%%%%%%%%%%%%%%%%%%%%%%%%%%%%%%%%%%%%%%%%%%%%%%%%%%%%%%%%%%%%%%%%%%%%%%%%%%%%%%%%
%% bibliography: see CFP for number of permitted pages

% \bibliographystyle{ACM-Reference-Format}  % do not change this line!
% \bibliography{bib/bib}  % put name of your .bib file here
\input{main.bbl}

\appendix
\section{Levy Flight vs. Normal Walk}
\input{doc/app1.tex}
\end{document}

%% file: doc/intro.tex
Inspiration for distributed systems can often be found in nature. Species like the cockroach, the \textit{Apis mellifera} honeybee, and the \textit{Temnothorax albipennis} rock ant are ideal examples of agents who interact locally to create emergent behaviours that benefit their entire swarm, hive, or nest \cite{cockroach-shelter-seeking,cockroach-site-selection-2, bee-quorum, ant-quorum}. These systems are studied both to correctly be able to model natural behavior and to modify it for the uses of engineered distributed systems like robotic swarms\cite{n-site-selection-summary}. \\
The house hunting behavior of the \textit{T. albipennis} rock ant has been thoroughly studied for their ability to quickly and successfully migrate their colony to a new nest of high quality after the old one has been destroyed. Many models try to replicate this behavior by modelling ants as probabilistic state machines and candidate sites as sites of different qualities that are equidistant from the home nest \cite{reina-favoring-model, pratt-paper-1, jiajia-paper}. These models are helpful in both autonomous swarm algorithms for tasks such as shelter seeking, and in biological research, to be able to better understand the ants' behavior. \\
One of the biggest faults of existing models and most empirical research, however, is that they either only experiment on sites that are equidistant from the nest, or they make the inherent assumption that all sites are equally likely to be found. This is not realistically true, as sites can be distributed anywhere and some sites may be harder to find than others due to obstacles like rivers or tall grasses. Furthermore, realistic representations of the difficulty in getting from one site to another are very influential in the actual outcome of a house-hunting decision -- it is important for models to be able to represent the fact that ants can choose a high quality candidate site even when it is much farther than other low quality sites, not just when all candidate sites are on a level playing field. We hope to open house-hunting modeling and empirical research to exploring the effects of the spatial orientation of candidate nest sites.\\
Our paper proposes augmentations to the ant house-hunting model in \cite{jiajia-paper} that allow the model to react to the spatial orientation of sites, resulting in higher granularity model parameters that can more accurately predict situations where the location of new candidate nests is relevant. Our model aims to both be representative of real ant's behavior, and to provide insights on the effect of different spatial configurations of nests in order to drive further empirical research on what real ants do in the same situations. We show that our model behaves similarly to real ants facing the same configurations.\\
The remainder of this paper is structured as follows. Section 2 describes the biological behaviours of the \textit{T. albipennis} rock ant, followed by a summary of the models in  \cite{jiajia-paper} and \cite{pratt-paper-1}, as well as the limitations that we seek to address. Section 3 describes the augmentations we have made to the model in \cite{jiajia-paper} in order for it to take into account the spatial orientation of nests during house hunting. Section 4 describes the specific parameters that we ran the model with and what information we were seeking to collect. Section 5 describes our results. In section 6, we further discuss the implications of our results, our conclusions, and areas of future research.

%% file: doc/background.tex
%background literature, including jiajia and pratt's work
\subsection{Biological Behavior}
When their old nest is destroyed, \textit{Temnothorax albipennis} ants are able to quickly search for a new nest and transport their colony to it within 24 hours. To do so, part of the colony designated as active ants participate in the following decision making process. The ants first leave their nest to search for new potential nest sites. Upon finding a candidate nest site, they examine the site's value and then return to their home nest, waiting a period of time inversely proportional to the new site's quality before recruiting other active ants and bringing them to the new site in a process known as forward tandem runs (FTRs). FTRs allow more ants to learn the path to the new site in case the colony decides to migrate there \cite{ant-quorum}. Once any ant encounters other ants at a threshold rate known as a \textit{quorum threshold} in a candidate nest site, they switch their behavior to a carrying behavior, where they carry other ants from the home nest to the candidate nest. During this phase, non-active ants (passive ants and brood items) get transported to the new home nest. The carrying process is three times faster than FTRs. This strategy enables the ants to new sites quickly and efficiently, accelerating the movement to the new nest when critical mass has been reached in it \cite{pratt-paper-1}. It also allows them to successfully choose higher quality nests over lower quality ones.   
\subsection{Existing models}
Existing models frequently represent each ant as a probabilistic state machine that is in one of four main phases -- Exploration and Assessment (sometimes together called the Uncommitted phase), Canvassing (also called the Favoring phase), and Transport (also called the Quorum or Committed phase). In the Exploration phase, agents are searching for new potential nest sites and do not yet have a preference. In the Assessment phase, agents have discovered a potential nest site and are evaluating it's favorability. In the Canvassing phase, agents have decided to support a particular net site and recruit others to try and convert them to have the same opinion (via tandem running). The last phase, the Transport phase, is only entered when a quorum (a locally high concentration of other agents) is sensed. In the Transport phase, ants who have committed to a particular new nest site begin carrying other members of their nest to the new site, and eventually the new nest becomes the ants' home nest. The full details of these models can be found in \cite{jiajia-paper, pratt-paper-1, cody-adams, reina-favoring-model} and many others. \\
We now go into further detail on the model in \cite{jiajia-paper}, which serves as the basis upon which we make distance-dependent modifications. In each of the four phases of this model, an agent can be in different types of states, which use empirically determined transition probabilities to transition into different states and phases. In the exploration ($E$) phase, ants can be in the \textit{at-nest}, \textit{search}, \textit{follow}, or \textit{arrive} states. Ants begin in the \textit{at-nest} state at their home nest, which has quality $0$. Nests in the model have a quality ranging from $0$ to $4$. With probability $1-p(x)$ where $x$ is the current nest quality and $p(x) = \frac{1}{1+e^{-\lambda x}}$ (where $\lambda$ is a model parameter), the ant transitions to the search state, where it begins to look for a new nest. With probability \textit{search-find}, the ant finds a new nest, where each nest other than the home nest is equally likely to be found, and transitions to the \textit{arrive} state. It is also possible for an ant in the \textit{at-nest} state to be led by another ant to a new nest, and in this case they successfully follow the other ant with probability \textit{follow-find}. Upon entering the \textit{arrive} state, ants can reject the new net with probability $1-p(x)$ or accept it with probability $p(x)$, in which case they enter the \textit{at-nest} state of the Assessment phase.\\
In the Assessment (A) phase, the ant has an equal probability of accepting their current nest or searching for another one, where search behavior as well as the \textit{search}-\textit{find}, \textit{follow-find}, and arrival probabilities are the same as they were in the Exploration phase. If the ant accepts the current nest, they transition to the $at\mhy nest$ state of the Canvassing phase.\\
In the Canvassing phase, the ant begins to actively recruit for their new candidate nest. From this phase, the ant can either enter the $quorum\mhy sensing$ state or the $search$ state, the latter of which remains with the same functionality as the $E$ and $A$ phases. In the $quorum\mhy sensing$ state, the ant senses a quorum if the current population of the nest is greater than the quorum threshold, and then enters the state $transport$ of the Transport phase. Otherwise, the ant enters the $lead\mhy forward$ state, which is representative of a forward tandem run. If the ant loses contact with the ant it is leading (which as stated previously, happens with probability $1-follow\mhy find$), the ant enters the $search$ state. Otherwise, the ant can continue to lead other ants forward. \\
Lastly, ants in the Transport phase, from the $transport$ state, can enter the $reverse\mhy lead$ state, which represents performing reverse tandem runs and leading other ants to their nest of choice. Ants can also enter the $search$ state, upon which the $arrive$ and $at\mhy nest$ states work the same as previously described.\\
In addition to these transitions, a few termination conditions are enforced to help ants decide on a new nest. For ants in the $transport$ state, if they keep trying to carry other ants in the $transport$ state, they will transition to the $at\mhy nest$ state of the Exploration phase with their new committed nest as the home net, since this suggests that most ants have already transitioned to the new nest. A similar condition exists for ants in the $lead\mhy forward$ state of the Canvassing phase, who also transition to the $at\mhy nest$ state of the Exploration phase if they try to lead many other ants who are also in the $lead\mhy forward$ state. 

%% file: doc/methodology.tex
The model in \cite{jiajia-paper} and many others, when modelling the \textit{search-find} probability, give each agent an equal probability of finding all of the non-home nest candidates. Models that simulate the ants travelling through space also typically only conduct experiments in situations where all candidate nests have an equal distance from the home nest. This is an important area of modelling work that has been neglected partially due to the shortage of data from real ant colonies when it comes to nests of differing qualities at different distances from the home nest. In order to augment the model in \cite{jiajia-paper}, we make the following changes. \\ 
\subsection{Modelling Sites as a Graph}
In our new model, the $n$ sites were represented as a complete undirected graph $G = (V, E)$ with edge weights representing the difficulty of traversing from one nest to the other. The ant's original home nest is denoted as node $v_0$. For each nest $v$, when an ant successfully found a nest from the $search$ state, the probability of which of the $n-1$ other nests $v'$ it found was represented by the function $\phi(v, v')$. \\
We provide two different ways to attain the function $\phi(v,v')$. The first methodology is analytical and more general, while the second one is specific to when edge weights represent distances in $\mathbb{R}^d$ or some other simulatable metric space.\\
%In order to be most accurate to actual ant behaviour, the function $\phi(v,v')$ for any given graph $G$ was computed via simulation, with c%
In the first methodology, $\phi(v,v')$ is defined as follows, where $e(v,v')$ represents the edge weight from node $v$ to node $v'$:
\begin{gather}
    \phi(v, v') = \frac{\frac{1}{e(v,v')}}{\sum\limits_{v'' \neq v} \frac{1}{e(v, v'')}}
\end{gather}
Note that as needed:
\begin{gather}
    \sum\limits_{v'\neq v}^{} \phi(v, v') = 1
\end{gather}
Here, the probability that a nest $v'$ is discovered is inversely weighted by the edge weight that represents the difficulty of discovery, since nests $v'$ with high discovery difficulty from $v$ are less likely to be found in a search. Other functions $\phi$ that satisfy equation $(2)$ may be used as well. Note also that instead of assigning each nest a location in some fixed metric space, edge distances were used instead to represent the graph between sites. This is because while metric space-based distances will result in a more precise graph structure that follows the triangle inequality, we may want a more flexible view of edge weights sometimes -- for example, nests in real life can be difficult to access if they are surrounded by a tall wall or or difficult-to-navigate terrain, like rivers and tall grass. Furthermore, the difficulty of access may depend on which other nest an ant is coming from. By using edge weights, we are able to capture these new situations by giving edges higher weight. Note that in the case where there is only $1$ nest, this model is equivalent to the model in \cite{jiajia-paper}, since only relative distances between two or more nests change the model behavior. Additionally note that when all pairwise distances between nests are scaled by a constant factor, model behavior does not change. 
\subsection{Distance-based Changes}
However, in the case where the graph can be represented as points in $\mathbb{R}^2$ or some other metric space, we can use a more accurate simulation-based methodology for calculating $\phi(v, v')$.\\
Given a specific graph $G$ of candidate nests, we build a separate simulator that takes in $G$, lays out the candidate nests in space, and calculates $\phi(v,v')$ by starting an ant at location $v$ and simulating the ants' movement via random walk until it is within sensing radius of another site $x \neq v$. This is done for a large number of rounds $R$, and then $\phi(v,v')$ is calculated as $\frac{R_{v'}}{R}$, where $R_{v'}$ is the number of rounds where the ant discovered site $v'$. (Note that $R_{v} = 0$ when we start our walk at nest $v$). \\
The random walk used in the simulation is the Levy flight random walk, which has been shown to be the preferred pattern over normal Brownian motion for animals to forage in  \cite{levy} when resources are sparse in the environment (as candidate nest sites are). More justification for the Levy flight can be found in Appendix A. \\

The simulator takes in parameters for the ants' speed $s$ and vision radius (which affects how easy it is to detect nearby sites), as well as $(\mu, c)$, the respective location and scale parameters for the Levy distribution used in the Levy flight random walk. Because the Levy distribution can generate extremely large values, we also included a $levy-cap$ parameter that capped the largest value generated so that ants would not walk too long in the same direction. \\

This methodology for finding transition probabilities is more accurate but must be separately calculated for each new nest configuration $G$, as opposed to the simpler estimates made in section 3.1. It is more computationally intensive and less general (only accounting for edge weights that represent distance) but is truer to the ants' actual behavior distance-based graphs, which are a very important use case for this model.\\

In order to properly reflect distance, we make further adjustments to the model. One important change that we need to make in this case is distance scaling, based off of the intuition that if an ant is in a nest $v$ with a higher average distance from all other nests $v'$, it will be harder for the ant to find another nest in the first place (corresponding to the \textit{search-find} probability in \cite{jiajia-paper}). Likewise, if an ant following another ant via forward tandem run to a new nest has to travel a longer distance, it is more likely that they will get lost (corresponding to $1-follow\mhy find$ in \cite{jiajia-paper}). Therefore, we augment \textit{search-find} as follows, where $f=search\mhy find$:
\begin{gather}
    search\mhy find'(v) = \begin{cases}
    \frac{a^2}{a_v^2}f & a_v > a\\
    search\mhy find & \text{otherwise}
    \end{cases}\\
    a = \frac{\sum\limits_{v' \neq v_0} e(v_0, v')}{n-1}\\
    a_v = \frac{\sum\limits_{v' \neq v} e(v, v')}{n-1}
\end{gather}
Note that $a$ is the average distance of other nests from the original home nest and $a_v$ is the average distance from all other nests from current nest $v$. Note that in this model we assume that the average distance of other nests to the home nest is a reasonable distance for the original \textit{search-find} parameter to still hold, but scale the ease of finding farther nests down with an inverse squared weighting factor (since area has a squared relationship to distance), so that:
\begin{gather}
    \lim\limits_{a_v \to \infty} search\mhy find'(v) = 0
\end{gather}
Similarly, for $follow\mhy find$, where an ant that is currently following another ant from nest $v$ to $v'$, we change this parameter to $follow\mhy find'(v, v')$, which is defined as follows, where $g = follow\mhy find$ and $a$ is defined in equation (4):
\begin{gather}
    follow\mhy find'(v, v') = \begin{cases}
        \frac{a^2}{e(v, v')^2}g & e(v, v') > a\\
        follow\mhy find & \text{otherwise}
    \end{cases}
\end{gather}
Here, we downweight the probability of following successfully (and increase the probability of getting lost) when the edge weight from $v$ to $v'$ has higher distance than $a$. 
% \subsection{Simulation-based Search Probabilities}
% In order to more accurately reflect the probability of finding a nest from any other nest, the model may be adapted into using simulation-based $\phi(v,v')$ probabilities in place of the probabilities mentioned in section 3.1. Given some specific layout of candidate nests and their pairwise distances, we built a separate simulator that takes in this layout and simulates the Levy flight random walk that ants in nature use. (Ants have been shown to favor the Levy flight over Brownian Motion random walks when resources are sparse like nest sites). The simulator takes in parameters for the ants' speed and vision radius, as well as $(\mu, c)$, the respective location and scale parameters for the Levy distribution used in the random walk. 500 random walks are run from each nest in the map, and end when some other site in the arena is discovered. Then, $\phi(v,v')$ is just the proportion of times random walks from $v$ ended up finding $v'$. \\
% This methodology of finding transition probabilities is more accurate but must be separately calculated for each new nest site configuration as opposed to the simplier estimates that are made in section 3.1.

\subsection{Nest-Quality-Dependent Parameters}\label{sssec:nqd} In actual house-hunting ant-colonies, the ants' tandem-running and quorum sensing behavior is dependent on site quality. Specifically, nests with a higher quality have a higher \textit{lead-forward} probability and a lower quorum threshold, which helps ants succeed in choosing higher quality but further away nests over lower quality but closer nests, even when the higher quality nest is up to nine times further than the low quality one \cite{distance-data}. These quality-dependent parameters can be be easily added to the model by setting \textit{lead-forward }to a function that scales up with higher quality, and \textit{QUORUM-THRE} (the quorum threshold) to a function that scales down with higher nest quality.  \\
To best match empirical results, these nest-quality dependent parameters are needed, as by taking distance into account we now disadvantage nests further away from the home nest by making them less likely to be found. In order to balance the trade-off between high quality but far away nests compared to closer and lower-quality nests, we must compensate by scaling \textit{lead-forward} and \textit{QUORUM-THRE} to favor higher quality sites, as intuitively makes sense. 

\subsection{Optional Specialized Changes} 
\subsubsection{Absolute Distances} Both the general and distance-based versions of our model by default assume relative distances in the calculation of \textit{search-find} and \textit{follow-find}. This is because scaling with distances is generally arbitrary for practical purposes and we assume that the average distance $a$ from candidate nests to the original nest corresponds to the \textit{search-find} and \textit{follow-find} probabilities in \cite{jiajia-paper}. However, sometimes it is important to examine the effect of different absolute distances, for example in experiments by O'Shea-Wheller et. al. \cite{real-distance}. To do so, we can add an extra hyperparameter $\alpha$ to the model easily, so that instead of using $a$, we use $\alpha$ to correspond to the original parameters in \cite{jiajia-paper}. This is particularly useful in the case of single-nest emigration, where the relative distance model simply assumes the parameter in \cite{jiajia-paper} no matter what distance to the single nest is specified. Our new updates now become:
\begin{gather}
    search\mhy find'(v) = \begin{cases}
    \frac{\alpha^2}{a_v^2}f & a_v > \alpha\\
    search-find & \text{otherwise}
    \end{cases}\\
    follow\text{-}find'(v, v') = \begin{cases}
        \frac{\alpha^2}{e(v, v')^2}g & e(v, v') > \alpha\\
        follow\text{-}find & \text{otherwise}
    \end{cases}
\end{gather}
While we did not implement this exact distance scale, we did implement a variation of it.
\subsubsection{Absolute Distance Implementation}\label{subsubsection:real_dist_implement} 
\begin{figure}
    \centering
    \includegraphics[width=0.45\textwidth]{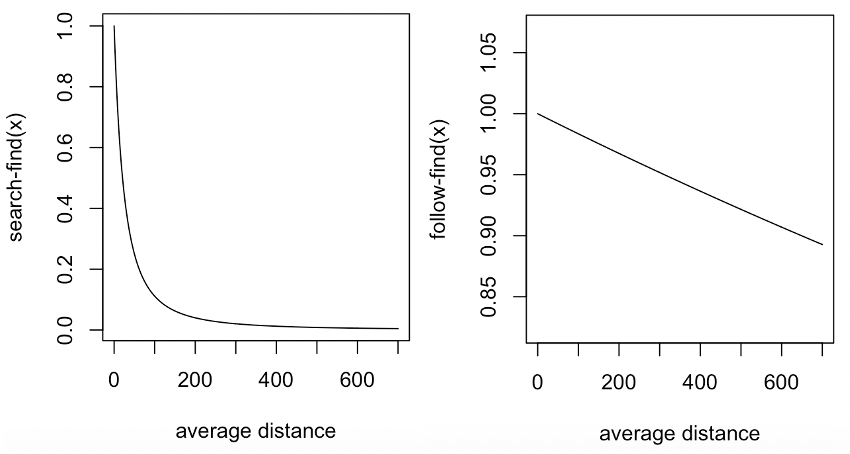}
    \caption{Values taken by $search\mhy find'$ and $follow\mhy find'$ as a function of distance, with the respective parameters $s=50$ and $t=12000$.}
    \label{fig:realdist_models}
\end{figure}

We implemented a variation of our absolute distance model, under the assumptions that (1) at distance 0, success rate should be 1; (2) rate decreases proportional to the inverse of the square of distance; (3) distance is scaled in millimeters. Let the parameters for $search\mhy find'$ and $follow\text{-}find'$ be $s$ and $t$ respectively. Then we have: 
\begin{gather}
    search\mhy find'(v) = \frac{s^2}{(a_v + s)^2}\\
    follow\text{-}find'(v, v') = \frac{t^2}{(e(v, v') + t)^2}
\end{gather}
Using a combination of results from \cite{jiajia-paper} and \cite{pratt-paper-1}, we approximate $s$ and $t$ under the assumption that $search\mhy find'$ should be 0.005 and $follow\text{-}find'$ should be 0.9 for a distance of 650 mm. We estimate $s=50$ and $t=12000$, with the corresponding $search\mhy find'$ and $follow\text{-}find'$ functions show in Figure \ref{fig:realdist_models}.

%% file: doc/setup.tex
\begin{figure}
    \centering
    \includegraphics[width=0.4\textwidth]{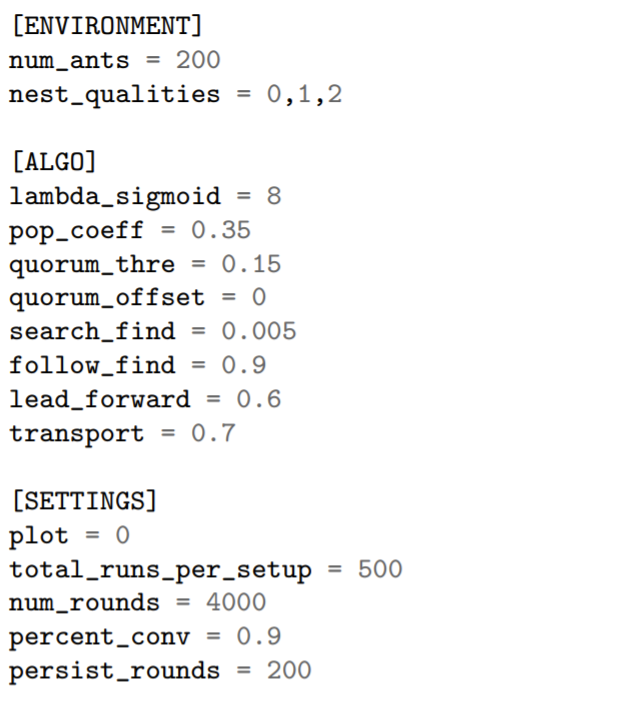}
    \caption{Original configuration parameters in \cite{jiajia-paper}. Additional graph configuration was added to support our new model.}
    \label{fig:params}
\end{figure}

In order to examine the accuracy of our model in comparison to real ants and to learn more about how distances affect the model itself, we kept most of the original parameters in \cite{jiajia-paper} the same while adding a \textit{graph-edges} parameter to the environment and changing \textit{num-rounds} to 2000 instead of 4000. In addition, for distance-based experiments, we also added in the $\phi(v,v')$ function, obtained via simulation as described in section 3.2. The original parameters in \cite{jiajia-paper} are shown in Figure \ref{fig:params}. Any additional changes to parameters are specified in the corresponding results section.

%% file: doc/results.tex
\subsection{Further Nest of Higher Quality}
\begin{figure}
    \centering
    \includegraphics[width=0.45\textwidth]{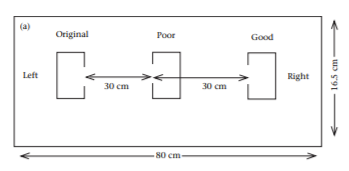}
    \includegraphics[width=0.45\textwidth]{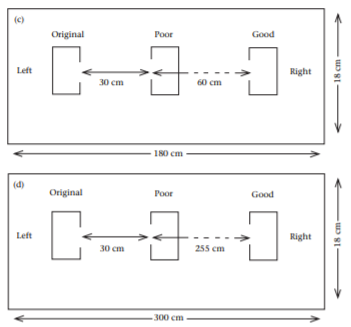}
    \caption{Experimental setup by Franks et. al. to test ants' ability to choose a further nest of higher quality over a closer, worse quality one.}
    \label{fig:distancetest}
\end{figure}

One of the most important applications of a model being able to take distance into account is it's ability to represent the trade-offs ants make between nest quality and distance from the home nest. House-hunting ants are capable of choosing high quality sites up to nine times further than a poor quality site, as shown by Franks et. al. \cite{distance-data}. However, when there are two equal quality sites, one close and one far, the ants efficiently choose the closer site. To test our model's ability to replicate this behavior, we copied the empirical setup of \cite{distance-data}, which can be seen in figure \ref{fig:distancetest}. In this setup, 15 different colonies were tested on their ability to migrate to a high quality nest when a low quality, closer nest was in the way. Three different distance comparisons were tested -- one where the better nest was twice as far as the worse one, one where the better nest was three times as far, and one where the better nest was nine times as far. Lastly, a control was tested where one nest was twice as far as the other but both were of equal quality. \\
In order to simulate this setup, we used a home nest quality of $0$, gave the closer nest a quality of $1$, and the further nest a quality of $2$, and ran each experiment for 500 trials. We tested two versions of our model -- one using the analytical $\phi(v,v')$ described in section 3.1, and the other using the simulated $\phi(v,v')$ described in section 3.2. The simulation parameters used in the second model version were as follows: we let $(\mu=10 \text{cm}, c=10)$ in the Levy flight along with an ant speed of $1$ cm per simulation time unit and a $levy-cap$ of $25$ time units (meaning ants could only walk in the same direction for at most $25$ time units). The simulated ants had a vision radius of $1.3$ cm. These parameters were obtained over trial and error until they best fit empirical results. More intuition behind the tuning of these parameters can be found in section \ref{subsec: walk sim tradeoffs}. \\
We used nest-quality dependent parameters as described in Section \ref{sssec:nqd}, where \textit{lead-forward} and \textit{QUORUM-THRE} were as follows for a candidate nest of quality $x$, and \textit{follow-find} was probability 0.5:\\
\begin{gather}
    lead\text{-}forward(x) = \begin{cases}
    0.9 & x = 4\\
    0.8 & x >= 3\\
    0.7 & x >= 2\\
    0.6 & x >= 1
    \end{cases}\\
    QUORUM\text{-}THRE(x) = \begin{cases}
    0.09 & x = 4\\
    0.11 & x >= 3\\
    0.13 & x >= 2\\
    0.15 & x >= 1
    \end{cases}
\end{gather}

\subsubsection{Model Performance}

The results of our two models on Franks' data, averaged over 100 trials, can be seen in Table \ref{tab:distance_results}. Since Franks' did not provide a description of the composition of their ant colonies, we ran our experiments on one of Pratt's ant colonies \cite{pratt-paper-1}, which contained 244 ants, 59 of them active, 74 of them passive, and 111 of them broods. The performance of the simulation based model can be seen in Table \ref{tab:distance_results_2}, while the analytical $\phi(v,v')$ model results can be seen in Table \ref{tab:distance_results}. As seen the both tables, there was no statistically significant difference between both model's performance and the performance of the actual ants, and our models succeeded in choosing the closer nest when both nests were of equal quality, but the farther nest when it was of higher quality. However, based on observation, the simulation-based model matched empirical results much more successfully than the analytical model, especially in the case where the high quality nest high was nine times as far. The simulation-based model only performed worse than the analytical model in the control case, because the simulation produced probabilities that were less extremely weighted towards the close nest, making the further but equally low quality nest slightly more likely to be found.  \\
Here, as in \cite{jiajia-paper} and \cite{pratt-paper-1}, we define the P-value as the proportion of simulations departing as far or farther from the colony average as did the experimental value, and only consider P-values of 0.05 or less to be statistically significant. 
\begin{table}[h]
    \centering
    \begin{tabular}{c|c|c|c|c|c}
         Quality & \makecell{Close \\ Dist} & \makecell{Far \\ Dist} & \makecell{Model \\ \%} & \makecell{Ants\\ \%} & P \\
         \hline\hline
         1, 4 & 30 & 60 & 91.6 $\pm$ 20.7 & 100 & 0.79 \\
         1, 4 & 30 & 90 & 81.9 $\pm$ 33 & 100 & 0.68 \\
         1, 4 & 30 & 285 & 52.3 $\pm$ 44 & 88 & 0.39 \\
         1, 1 & 30 & 60 & 92.3 $\pm$ 19 & 93  & 0.8 \\ 
    \end{tabular}
    \caption{Accuracy of model in choosing the highest quality rank, tiebreaking by closest distance. Note in the quality column the first number is the quality of the closer nest, and the second is the quality of the further nest. Distances are in cm.}
    \label{tab:distance_results}
\end{table}
\begin{table}[ht]
    \centering
    \begin{tabular}{c|c|c|c|c|c}
         Quality & \makecell{Close \\ Dist} & \makecell{Far \\ Dist} & \makecell{Model \\ \%} & \makecell{Ants\\ \%} & P \\
         \hline\hline
         1, 4 & 30 & 60 & 99.5 $\pm$ 1.2  & 100 & 0.8 \\
         1, 4 & 30 & 90 & 95.7 $\pm$ 18.7 & 100 & 0.74 \\
         1, 4 & 30 & 285 & 83.5 $\pm$ 35.7 & 88 & 0.68 \\
         1, 1 & 30 & 60 & 81.5 $\pm$ 38.2 & 93  & 0.82 \\ 
    \end{tabular}
    \caption{Accuracy of model using the more accurate simulation-based searching probabilities. Note in the quality column the first number is the quality of the closer nest, and the second is the quality of the further nest. Distances are in cm.}
    \label{tab:distance_results_2}
\end{table}
\subsubsection{Validation of Model Behavior}
\begin{figure}
    \centering
    \includegraphics[width=0.45\textwidth]{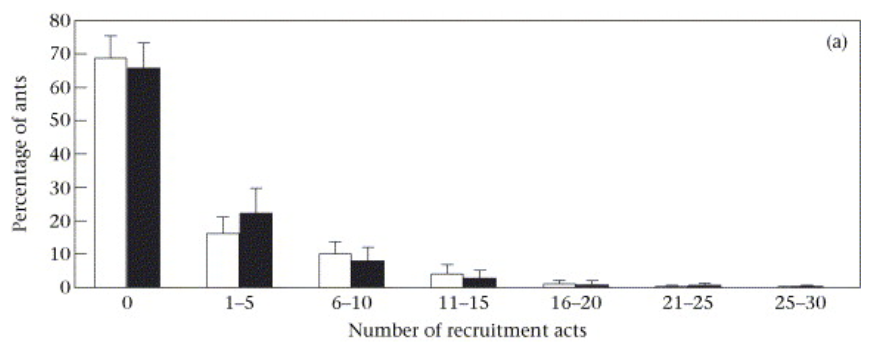}
    \includegraphics[width=0.45\textwidth]{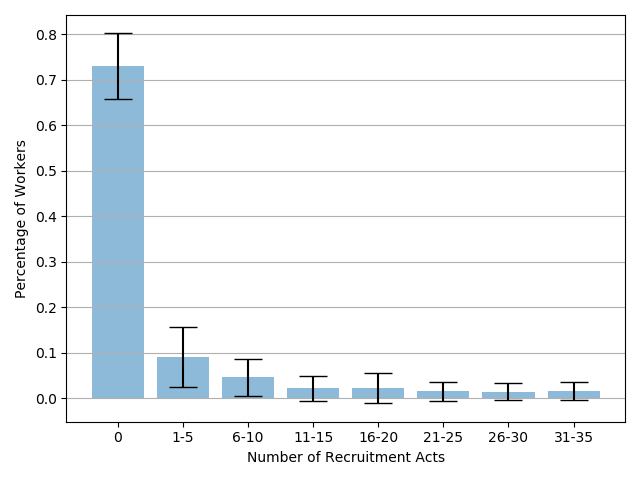}
    \caption{Pratt's empirical recruitment act results are shown by the black bars in the top graph.  Our figure was produced using the distances in setup (a) of Figure \ref{fig:distancetest}, with the close nest having quality 1, and the far nest with quality 4.}
    \label{fig:recruit}
\end{figure}

In order to validate that the remainder of our Levy Flight simulation-based model still matched actual ant behaviour in ways other than the ants' final nest choice, we compared the histogram of recruitment acts performed with the one observed in Pratt's empirical results \cite{pratt-paper-1}. We observe a similar pattern in the recruitment acts, which appear to be somewhat exponentially decreasing, with a very large number of acts making 0 recruitment acts (Figure \ref{fig:recruit}). 

\subsection{Trade-offs created in simulation parameters} \label{subsec: walk sim tradeoffs}
In order to tune the $\phi(v,v')$ to match empirical results, the following trade-offs were observed in our Levy Flight model and helped to guide our trial-and-error approach. 
\subsubsection{Vision Radius}
\begin{figure}
    \centering
    \includegraphics[width=0.45\textwidth]{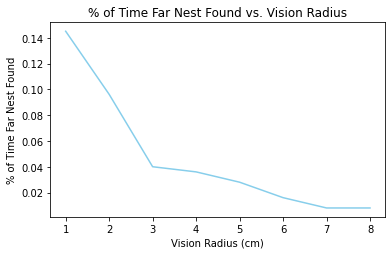}
    \includegraphics[width=0.45\textwidth]{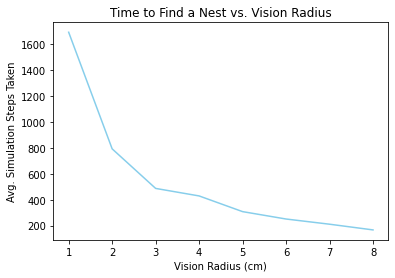}
    \caption{Nest search time and percentage of times the far nest was found compared to vision radius. All other parameters in the simulation other than vision radius were kept the same as in section 5.1.}
    \label{fig:rad}
\end{figure}

A larger vision radius helped ants find a nest $v'$ from $v$ quicker in the random walk simulations, but also made it less likely that further nests would be found. Figure \ref{fig:rad} shows the nest search time as well as the percentage of times the further nest was found for varying vision radii, run using the distances in setup (c) of Figure \ref{fig:distancetest} (where the higher quality site was more than 9 times further than the low quality site). The percentage of times the far nest was found compared to vision radius is relevant in tuning the distance-quality tradeoff, though we must keep vision radius low realistically to be representative of ants. 
\subsubsection{Levy Cap vs. Decision Accuracy}
\begin{figure}
    \centering
    \includegraphics[width=0.45\textwidth]{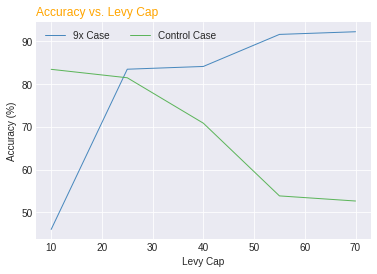}
    \caption{Tradeoff between accuracy on the control case and accuracy on the nine times as far case in section 5.1. As the levy cap increases, the control accuracy gets worse (since the two sites become undistinguishable), and the non-control accuracy improves.}
    \label{fig:levy-cap}
\end{figure}

In tuning $\phi(v,v')$ an inherent tension was discovered between accuracy in the control case of section 5.1, and accuracy in the non-control cases. Specifically, a larger Levy Cap (allowing longer step sizes to be drawn from the distribution, making the random walk more extreme in nature) made the further nest easier to find compared to the closer nest. The easier it is to find the further nest, the more accurate the house hunting model will be when the further nest is better than the closer one, but the less accurate it will be when the further nest is of equal quality to the closer one. This trade-off can be observed in Figure \ref{fig:levy-cap}.

\subsection{Two Nests Equally Distant from Home}\label{subsec:twoequidist}
To validate that the model is still accurate to empirical \textit{Temnothorax} behavior, we ran both versions of our new model using the ant colonies described in Figure \ref{tab:two_site}, taken from Pratt's experiments in \cite{pratt-paper-1}. We used nest qualities of $1$ and $2$ and a home nest quality of $0$. The two candidate nests were a distance of 650 mm from the home nest and a distance of 280 mm from eachother, as in Pratt's original experiment. We used the same simulation parameters as in section 5.1 for the Levy Flight simulation model. As with Zhao's model, the resulting $p$-values indicate that there is no statistically significant difference between both of our models and the empirical results for the first five colonies. However, the simulation-based model matches the third, fourth, and fifth colony results very well but is more accurate than the actual ants for the other three colonies. Note that only active workers participate in the algorithm, and the rest of the colony is transported.
\begin{table}[h]
    \centering
    \begin{tabular}{c|c|c|c|c}
         (A,P,B) & Model S & Model A & Zhao & Ants  \\
         \hline \hline 
         (70,28,228) & \makecell{86 $\pm$ 18\\0.12}&\makecell{86 $\pm$ 22\\0.12} & \makecell{51 $\pm$ 17\\0.86} & 61 \\
         \hline 
         (59, 74, 111) & \makecell{92 $\pm$ 12\\0.12} &\makecell{89 $\pm$ 25\\0.2} & \makecell{61 $\pm$ 28\\0.56} & 80\\
         \hline 
         (62, 95, 106) & \makecell{98 $\pm$ 10\\0.94} &\makecell{90 $\pm$ 23\\0.77} & \makecell{63 $\pm$ 30 \\0.36} & 99\\
         \hline 
         (67, 42, 192) & \makecell{98 $\pm$ 19\\0.94} &\makecell{88 $\pm$ 23\\0.69} & \makecell{59 $\pm$ 20 \\ 0.1} & 98 \\
         \hline 
         (53, 88, 61) & \makecell{98 $\pm$ 9\\0.96} &\makecell{86 $\pm$ 30\\0.78}& \makecell{61 $\pm$ 34 \\ 0.5} & 100\\
         \hline 
         (73, 101, 173) & \makecell{94 $\pm$ 14\\0.0} &\makecell{92 $\pm$ 18\\0.01} & \makecell{60 $\pm$ 25 \\ 0.02} & 2\\
         
    \end{tabular}
    \caption{Percentage of brood items in higher quality nest when home nest becomes empty. P values are presented in the second row for each model. Model S is the Levy flight simulation based model, and Model A is the analytical model.}
    \label{tab:two_site}
\end{table}

\subsection{Real Distance}
\subsubsection{Two Nests Equidistant From Home But Closer To Each Other} To ensure that our absolute distance implementation produces comparable results to our regular analytical model in a situation where the average distance to the home nest is considered to match the parameters in \cite{jiajia-paper}, we validate our incorporation of real distance using the same experimental data and model parameter setup as in Section \ref{subsec:twoequidist}. The graph edges between home nest and candidate nests were set to 650, and the graph edge between the two candidates was set to 280, in line with Pratt's original experimental setup \cite{pratt-paper-1}. We use $s=50, t=12000$ for our real distance rate model's parameters, as described in \ref{subsubsection:real_dist_implement}. We ran 500 simulations for each ant colony composition detailed in Table \ref{tab:two_site_realdist} below. Again, the predictions for 5 of the colonies do not differ significantly from empirically observed values and from the analytical model, indicating that the real distance implementation is comparable and will represent ant behaviour in single-nest emigrations well. 
\begin{table}[h]
    \centering
    \begin{tabular}{c|c|c|c|c}
         \makecell{Act, Pass,\\ Brood} & \makecell{Real\\ Dist.} & \makecell{Model\\A} & Zhao & Actual  \\
         \hline \hline 
         70, 28, 228 & \makecell{85 $\pm$ 22\\0.13} & \makecell{86 $\pm$ 22\\0.12} & \makecell{51 $\pm$ 17\\0.86} & 61 \\
         \hline 
         59, 74, 111 & \makecell{81 $\pm$ 32\\0.29} &\makecell{89 $\pm$ 25\\0.2} & \makecell{61 $\pm$ 28\\0.56} & 80\\
         \hline 
         62, 95, 106 & \makecell{81 $\pm$ 32\\0.64} & \makecell{90 $\pm$ 23\\0.77} & \makecell{63 $\pm$ 30 \\0.36} & 99\\
         \hline 
         67, 42, 192 & \makecell{84 $\pm$ 24\\0.54} &\makecell{88 $\pm$ 23\\0.69} & \makecell{59 $\pm$ 20 \\ 0.1} & 98 \\
         \hline 
         53, 88, 61 & \makecell{81 $\pm$ 34\\0.69} & \makecell{86 $\pm$ 30\\0.78}& \makecell{61 $\pm$ 34 \\ 0.5} & 100\\
         \hline 
         73, 101, 173 & \makecell{83 $\pm$ 26\\0.02} &\makecell{92 $\pm$ 18\\0.01} & \makecell{60 $\pm$ 25 \\ 0.02} & 2\\
    \end{tabular}
    \caption{Percentage of brood items in higher quality nest when home nest becomes empty, compared between our model accounting for real distance, our relative distance model, Zhao's results, and Pratt's actual experimental results. }
    \label{tab:two_site_realdist}
\end{table}

\subsubsection{Single Superior Nest With Differing Distance to Home} 
\begin{figure}
    \centering
    \includegraphics[width=0.45\textwidth]{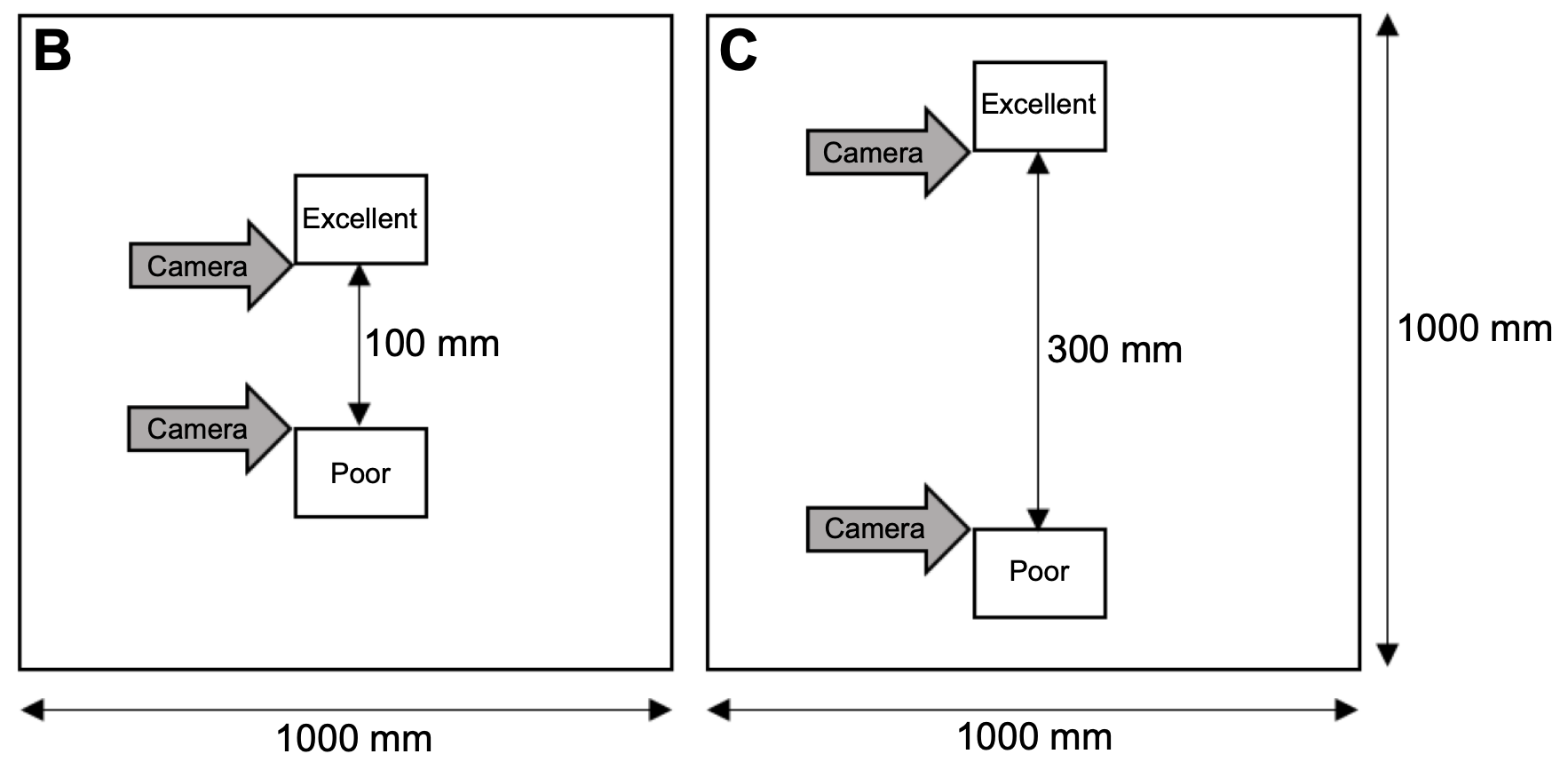}
    \caption{Experimental setup by O'Shea-Wheller et. al. to elucidate differences in ant behavior arising from differing distance to candidate nest site.}
    \label{fig:realdist_experiment}
\end{figure}

\begin{figure}
    \centering
    \includegraphics[width=0.45\textwidth]{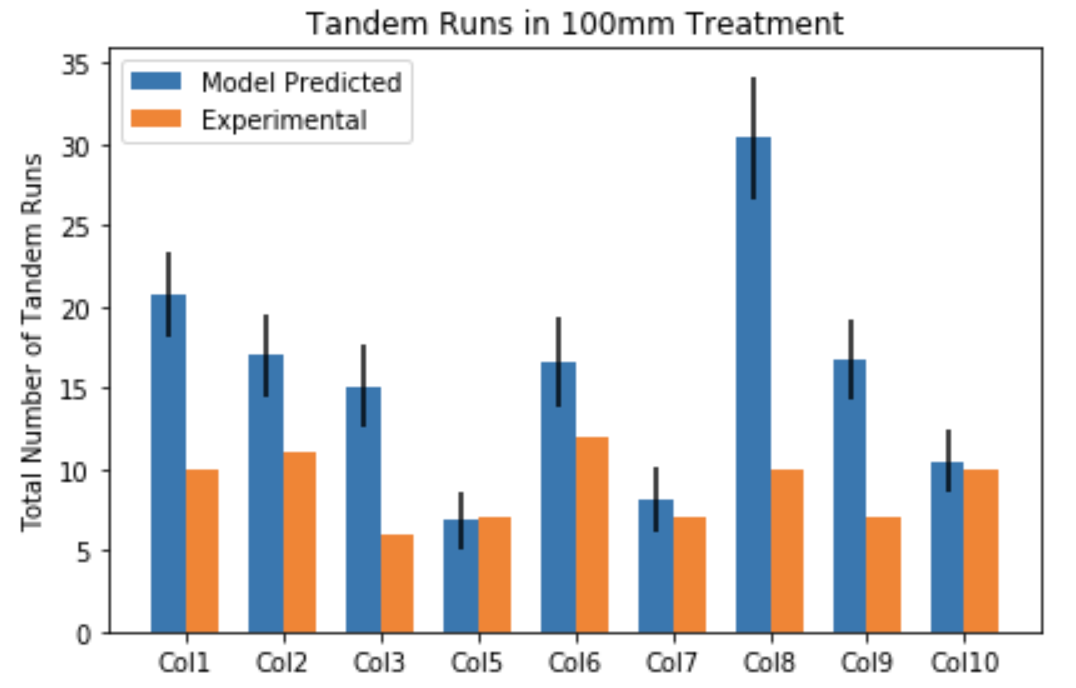}
    \includegraphics[width=0.45\textwidth]{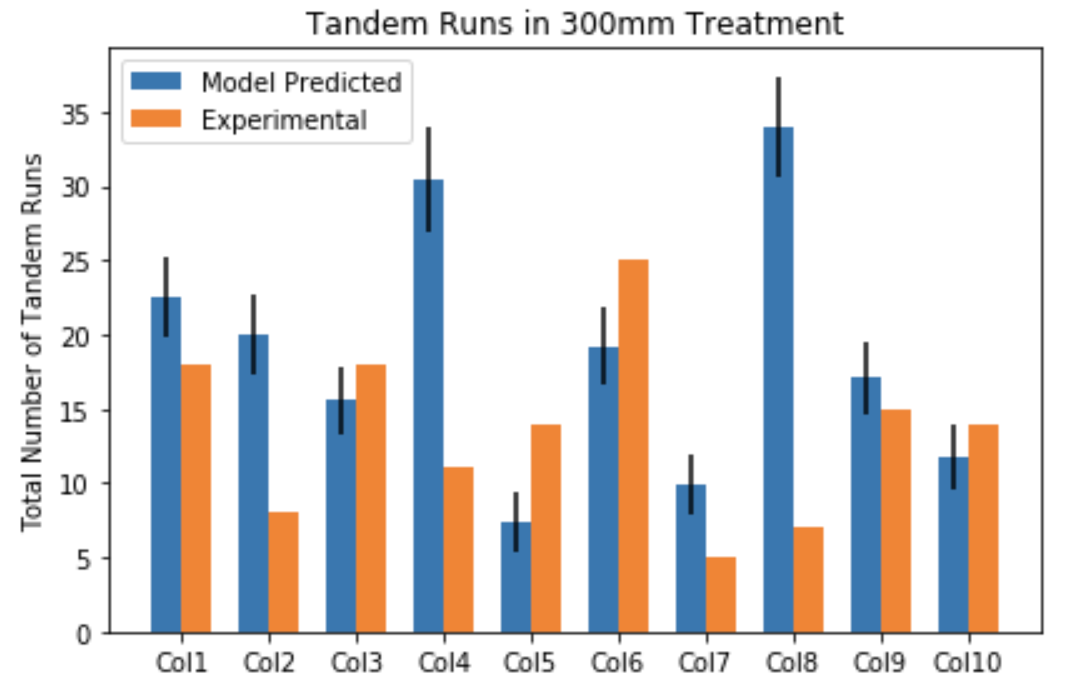}
    \caption{Comparison of total number of tandem runs predicted by model and empirically observed by O'Shea-Wheller et. al.}
    \label{fig:realdist_tandem}
\end{figure}

In order to truly assess our real distance implementation, we also validate against empirical data from a setup meant to test variations in ant behavior solely attributed to distances. O'Shea-Wheller et. al. observed 10 different ant colonies in both of two experimental situations where there was a single candidate nest of quality much higher than that of the home nest, providing incentive for the ants to move. In one treatment the distance between the two sites was 100 mm, and in the other the distance was 300 mm, as seen in Figure \ref{fig:realdist_experiment} \cite{real-distance}. O'Shea-Wheller et. al. tracked each ant and collected data including total number of tandem runs, time to first discovery of new nest site, and time to quorum at the new site.
To tailor simulation parameters to experimental conditions, home nest quality was set to 0.5, representing a relatively poor nest, and candidate nest quality was set to 2, representing an excellent nest. The distance between the two sites was 100 for the simulation corresponding to the 100 mm treatment, and 300 for simulating the 300 mm treatment. Colony compositions, which consists of number of workers in a colony, as well as number of active workers in each treatment (which differed between treatments for the same colony) was specified by O'Shea-Wheller for each of the colonies. While colony compositions were provided, there was only a maximum and minimum provided for number of brood items, so we estimated number of brood items proportional to relative position in the given range. That is, as range of number of brood items is [19, 130] and range of number of workers is [47, 187], our estimate was: $\text{brood-estimate} = \frac{\text{num-workers} - 47}{187 - 47} \cdot (130 - 19) + 19$. We ran 200 simulations for each of the 10 colonies, for both treatments, with results displayed in Figure \ref{fig:realdist_tandem}.

%% file: doc/discussion.tex
Results from our graph-based model employing relative distances for the two candidate nests with different distance align very well with empirical data. When the near and far candidate nests are of the same quality, our model predicts that the ants choose the near one, but if the far one is better in quality, our model predicts that they choose the far one. The simulation-based methodology proved, as predicted to give a more accurate match to empirical data. However, in two cases -- when the far nest is approximately 9 times the distance to the near nest, and the control case where the far and close nest had the same quality -- were in inherent tension in our model, 
as a more balanced probability of choosing each nest made it easier for the further and better nest to be found, but also made it harder for ants to prefer the close nest when both nests were of similar quality. This suggests that probabilities may not be enough on their own, and adding a direct perception of distance to the ants' value function could help with this tension. The analytical $\phi(v,v')$ version of the model did worse than actual ants at finding the far nest, especially in the case where the far nest was nine times further. This indicates that the analytical model prioritizes closeness over quality more than the ants do.\\

% The results for the effect of the measure of the convex hull on how badly the ants split and how long it takes the ants to move nests is mildly surprisingly in how amorphous the relationship appear to be. In fact, there is not any obvious correlation between the oblongness of the nearby nest positions and those other statistics at all, where it might have seemed initially that the more stretched out the nests are from each other, the more likely ants would be to split and the longer moving would take. This \textit{could} still be true, as all the data points are well within each others confidence intervals. In this case more simulations simply need to be run. However, there are also a few factors contributing to the noise and removing the relationship between oblongness and split-off proportion/number of rounds taken.

% The disparities in diameter for a given number of vertices is less impressive than it seems. This is because the original nest is chosen randomly from the vertices in the completed graph. Thus, a diameter of $4$ for a path graph of length $5$ could still means that the original nest is at the center and the average distance from another site to the original nest is $1.5$ edge units. Similar logic holds for the uniformly distributed nest locations.

Additionally, as demonstrated in Section 5.2, nests significantly farther away (even if only by a factor of 2) were demonstrably less likely to be picked by ants, so one of the more important factors was actually the distance of the closest nest to the original nest. In the case of the randomly-generated graphs, the distance to the closest nest was always 65, so results really just depended on how many other nest sites were nearby and how well the ants could resolve splits (which is well, as mentioned by most papers, including Zhao's). In the randomly-generated locations case, it would be even more likely that one nest is significantly closer than the others. At the very least, using both relative graph distances and real distances appear to provide the same general result that the exact distribution of possible nest sites outside the original ant colony does not appear to be very important.

Generally, for the experimental setup with two candidate nests of differing qualities equidistant to the home nest, our results improve on previous results. The graph-based model predicts greater favoring of the higher quality nest than the original model, which is generally appropriate when considering experimental data. The real-distance graph-based model predicts that ants favor the higher quality nest to a lesser extent than our relative-distance graph-based model, though still more than the original model, which also results in improvement in performance.

As for the real-distance specific simulations, in each of the two treatments, total number of tandem runs in approximately a third of the colonies are quite close to the actual observations. For all but two colonies in the 100 mm treatment and three colonies in the 300 mm treatment, model-predicted and actual results are not exceedingly different, though could still be improved. Total tandem runs are generally overestimated for 100 mm, while there is no clear visible trend for 300 mm. This is consistent with O'Shea-Wheller's findings, where tandem running attempts increase significantly for 300 mm \cite{real-distance}. Our model does not have an explicit mechanism to account for this, and thus could overestimate tandem running for 100 mm distance. While we do see that total number of tandem runs for 300 mm is larger, the difference is not particularly large. 

In addition, time to first scout discovery of new nest and time to quorum were measured, however, comparing to the O'Shea-Wheller's results proves difficult due to consistently very large standard deviations. We suspect that this is due to the nature of the discrete time steps in the execution of the simulation, especially as they don't correspond directly to real time.

%% file: doc/future.tex
Future work includes examining the effects of spatial configurations in autonomous swarm models designed to solve the house hunting problem such as \cite{reina-favoring-model} where agents are actually simulated moving through space, so it is not necessary to resort to trying to capture spatial distribution via probabilities of discovery. The trade-off with using simulations of this form is that the model can no longer capture difficulties in finding a site that do not have to do with distance, but is nonetheless still very important to explore. In addition, these simulations may prove more expensive. \\

In addition, future work needs to be done in empirical experiments to collect a greater number of results on how ants react to nest sites at different distances. Almost all existing experiments on house-hunting ants involve only one candidate nest or have candidate nests at the same distance. Gathering data on how nest distances affect ant behaviour will help validate new house-hunting models and give us greater insight into ant behavior so that we can improve the new models.\\

One particularly important experiment to run would be one that determines whether distance is intrinsically incorporated into ants' evaluation of a site's quality, or if the results observed in \cite{distance-data} (where ants will choose the closer nest out of two equal quality nests, but the further nest when that nest is better) arise naturally. Since ants wait a period of time inversely proportional to perceived site quality, one could test this by seeing how long the same ants wait before recruiting for a closer nest vs. a further nest of the same quality. If they wait different times, this would imply that distance is something the ants take into account when evaluating site quality. This information would be very useful for distance-based house hunting models, because it provides biological motivation for creating a site quality function which directly includes a distance term. Such a model would bring us closer to empirical results, but needs to be justified first. \\

% edit, wayne section
There is also much to be done in terms of studying situations with more than two candidate nests. We have started this process by examining the effects of graph diameter and other measures of nest distribution by randomly generating graphs with 3-6 candidate nests of similar quality, but have not yet shown a relationship between these distribution measures and the percentage of ants in the best nest or the time it takes ants to converge. Further analysis and experimentation is needed to understand the correlation between these variables.\\ 
In addition to randomly generated graphs, more specific graphs could be constructed to test this. For instance, one might compare rooted trees of nests as vertices with the same number of vertices but different heights, and with the roots at the original nest. It would also be interesting to purposefully split the ants initially and place them on different vertices of a graph and see how long it takes for the ants to merge back together. As mentioned, the ants appear to handle splits very well, which means data about what happens when there \textit{is} a split and how the nest location distributions affect that is relatively lacking.

Our real-distance implementation could possibly be improved by using a different rate function, and perhaps even relaxing some assumptions. For example, we might not necessarily insist that a distance of 0 should correspond to a rate of 1 (which we imposed because discovering a location that an agent is already at should be guaranteed), or use a function that decays faster than $\propto \text{distance}^{-2}$ (though that might sacrifice the area-based interpretation of search probability). Even keeping the current model, the parameters $s$ and $t$ could be tuned, especially using O'Shea-Wheller's observed tandem run success rates to tune \textit{follow-find}. Using a simulation of ants traveling, as mentioned above, might also yield improvements in this area. Another slightly related remark is that our current model still does not account for whether or not an ant has previously encountered a given nest site, which could improve the realism of the model. Finally, real distance does not integrate obstacle difficulty. This could possibly be done with an additional input parameter specifying obstacles on edges, though it will be difficult to ground the value in reality.

Further comparisons with O'Shea-Wheller's work could be made. For example, investigating percentage of colony involved in tandem runs could prove fruitful, though whether or not O'Shea-Wheller counted scouts among this number needs to be ascertained. Also, distribution of number of tandem runs per worker ant can be compared to empirical results. It might be interesting to pair single simulations of 100 mm and simulations of 300 mm and compare which worker does what in both simulations, which is something that O'Shea-Wheller focused on, but that might be difficult. First discovery time and quorum time had extremely high variance, so getting more reliable results for that might be good for comparison purposes. The current model assumes that all active ants scout, but empirical results suggest that that might not be the case, and number of scouts has dependency on distance. Incorporating this into the model might improve realism and performance. O'Shea-Wheller also considered worker tandem running return speeds being greater for the 300 mm treatment, which is possibly a conscious adjustment by the ants in response to greater distance. This could possibly be built in, in the future, though it appears difficult.

%% file: main.bbl

%% file: doc/app1.tex
\begin{figure}
    \centering
    \includegraphics[width=0.45\textwidth]{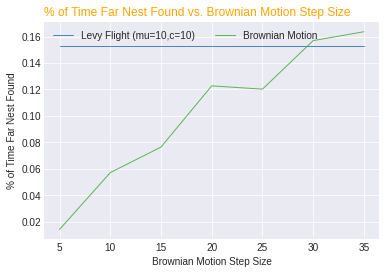}
    \caption{Percentage of times nine times as far site was visited for various Brownian motion step sizes compared to a Levy flight with $(\mu=10,c=10)$ and step size capped at 25.}
    \label{fig:brown_step_size_1}
\end{figure}
\begin{figure}
    \centering
    \includegraphics[width=0.45\textwidth]{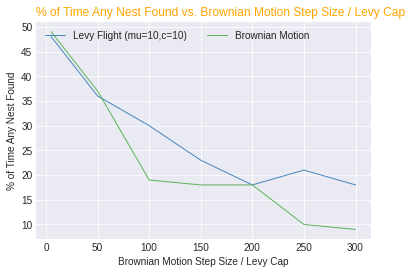}
    \caption{Percentage of times nine times as far site was visited for various Brownian motion step sizes compared to a Levy flight with $(\mu=10,c=10)$ and step size capped at the same step size.Sensing radius was set to $5$ cm.}
    \label{fig:brown_step_size_2}
\end{figure}

To further examine the results in \cite{levy} which stated that animals prefer the Levy flight while foraging and justify the use of Levy flight as opposed to Brownian motion in our model, we used our simulator to examine the effects of the walk type used on the ants' ability to find sites further from the nest. \\
\begin{figure}
    \centering
    \includegraphics[width=0.23\textwidth]{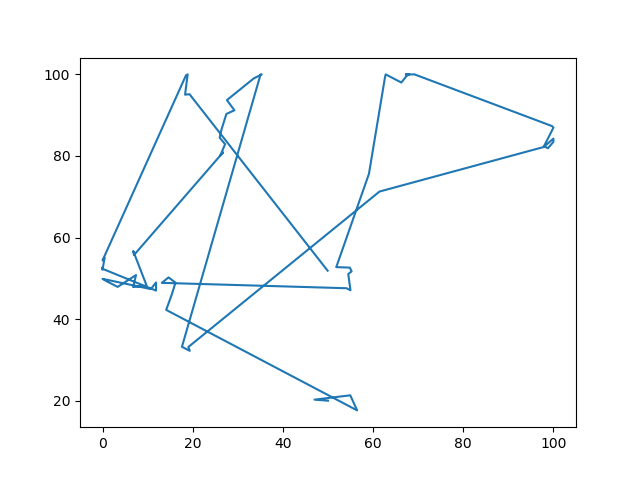}
    \includegraphics[width=0.23\textwidth]{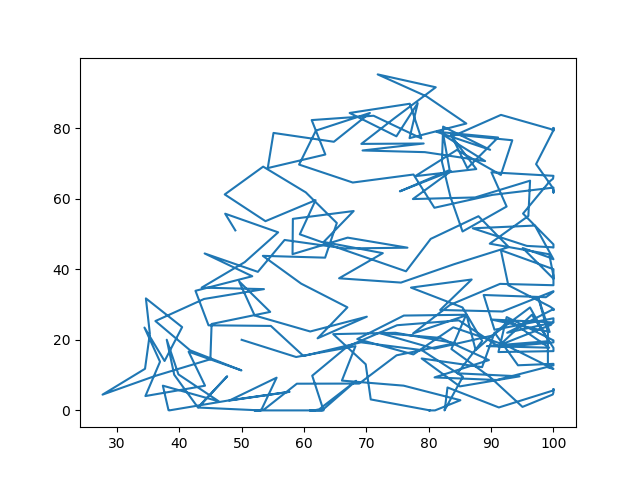}
    \caption{Levy flight pattern (left) vs Brownian motion (right) random walk pattern of a simulated ant searching for a new nest site. }
    \label{fig:my_label}
\end{figure}
Intuitively, the Levy flight is preferable because it both allows ants to make small localized explorations and longer jumps around the search space by sampling varying step sizes from the Levy distribution. Contrarily, the Brownian motion random walk simply uses the same step size, so ants always travel for the same distance before switching direction. Thus, if the Brownian motion step size is small, it is rare that ants will travel far enough to find further sites, but if the step size is too large, the ants will end up too far away from their home nest. \\
Figure \ref{fig:brown_step_size_1} shows the percentage of times a site $B$, nine times as far from the home nest as site $A$, is found compared to the percentage of times that site $A$ is found in a bounded arena, for the Levy flight with parameters $(\mu=10, c=10)$, capped at max step size $25$ compared to Brownian motion with various step sizes. Note that a smaller normal walk step size makes it harder for the further nest to be discovered, while a large step size of around $25$ or $30$ results in approximately the same chances of discovery.\\
Figure \ref{fig:brown_step_size_2} shows the motivation for using Levy flight as opposed to a Brownian motion walk with large step size -- in an unbounded arena, it is much less likely for any sites to be found when search time is capped.

%% file: main.bbl

\begin{thebibliography}{12}

%%% the Web address.

\ifx \showCODEN    \undefined \def \showCODEN     #1{\unskip}     \fi
\ifx \showDOI      \undefined \def \showDOI       #1{#1}\fi
\ifx \showISBNx    \undefined \def \showISBNx     #1{\unskip}     \fi
\ifx \showISBNxiii \undefined \def \showISBNxiii  #1{\unskip}     \fi
\ifx \showISSN     \undefined \def \showISSN      #1{\unskip}     \fi
\ifx \showLCCN     \undefined \def \showLCCN      #1{\unskip}     \fi
\ifx \shownote     \undefined \def \shownote      #1{#1}          \fi
\ifx \showarticletitle \undefined \def \showarticletitle #1{#1}   \fi
\ifx \showURL      \undefined \def \showURL       {\relax}        \fi
% invisible to TeX
\providecommand\bibfield[2]{#2}
\providecommand\bibinfo[2]{#2}
\providecommand\natexlab[1]{#1}
\providecommand\showeprint[2][]{arXiv:#2}

\bibitem[\protect\citeauthoryear{Cody and Adams}{Cody and Adams}{2017}]%
        {cody-adams}
\bibfield{author}{\bibinfo{person}{J.~R. Cody} {and} \bibinfo{person}{J.~A.
  Adams}.} \bibinfo{year}{2017}\natexlab{}.
\newblock \showarticletitle{An evaluation of quorum sensing mechanisms in
  collective value-sensitive site selection}.
\newblock  (\bibinfo{date}{Dec} \bibinfo{year}{2017}), \bibinfo{pages}{40--47}.
\newblock
\urldef\tempurl%
\url{https://doi.org/10.1109/MRS.2017.8250929}
\showDOI{\tempurl}


\bibitem[\protect\citeauthoryear{Franks, Hardcastle, Collins, Smith, Sullivan,
  Robinson, and Sendova-Franks}{Franks et~al\mbox{.}}{2008}]%
        {distance-data}
\bibfield{author}{\bibinfo{person}{Nigel~R. Franks},
  \bibinfo{person}{Katherine~A. Hardcastle}, \bibinfo{person}{Sophie Collins},
  \bibinfo{person}{Faith~D. Smith}, \bibinfo{person}{Kathryn~M.E. Sullivan},
  \bibinfo{person}{Elva~J.H. Robinson}, {and} \bibinfo{person}{Ana~B.
  Sendova-Franks}.} \bibinfo{year}{2008}\natexlab{}.
\newblock \showarticletitle{Can ant colonies choose a far-and-away better nest
  over an in-the-way poor one?}
\newblock \bibinfo{journal}{\emph{Animal Behaviour}} \bibinfo{volume}{76},
  \bibinfo{number}{2} (\bibinfo{year}{2008}), \bibinfo{pages}{323 -- 334}.
\newblock
\showISSN{0003-3472}
\urldef\tempurl%
\url{https://doi.org/10.1016/j.anbehav.2008.02.009}
\showDOI{\tempurl}


\bibitem[\protect\citeauthoryear{Garnier, Jost, Jeanson, Gautrais, Asadpour,
  Caprari, and Theraulaz}{Garnier et~al\mbox{.}}{2005}]%
        {cockroach-site-selection-2}
\bibfield{author}{\bibinfo{person}{S. Garnier}, \bibinfo{person}{C. Jost},
  \bibinfo{person}{R. Jeanson}, \bibinfo{person}{J. Gautrais},
  \bibinfo{person}{M. Asadpour}, \bibinfo{person}{G. Caprari}, {and}
  \bibinfo{person}{G. Theraulaz}.} \bibinfo{year}{2005}\natexlab{}.
\newblock \showarticletitle{Collective decision-making by a group of
  cockroach-like robots}. In \bibinfo{booktitle}{\emph{Proceedings 2005 IEEE
  Swarm Intelligence Symposium, 2005. SIS 2005.}} \bibinfo{pages}{233--240}.
\newblock
\urldef\tempurl%
\url{https://doi.org/10.1109/SIS.2005.1501627}
\showDOI{\tempurl}


\bibitem[\protect\citeauthoryear{O'Shea-Wheller, Wilson-Aggarwal, Edgley,
  Sendova-Franks, and Franks}{O'Shea-Wheller et~al\mbox{.}}{2016}]%
        {real-distance}
\bibfield{author}{\bibinfo{person}{Thomas~A. O'Shea-Wheller},
  \bibinfo{person}{Deraj~K. Wilson-Aggarwal}, \bibinfo{person}{Duncan~E.
  Edgley}, \bibinfo{person}{Ana~B. Sendova-Franks}, {and}
  \bibinfo{person}{Nigel~R. Franks}.} \bibinfo{year}{2016}\natexlab{}.
\newblock \showarticletitle{A social mechanism facilitates ant colony
  emigrations over different distances}.
\newblock \bibinfo{journal}{\emph{Journal of Experimental Biology}}
  \bibinfo{volume}{219}, \bibinfo{number}{21} (\bibinfo{year}{2016}),
  \bibinfo{pages}{3439 -- 3446}.
\newblock
\urldef\tempurl%
\url{https://doi.org/10.1242/jeb.145276}
\showDOI{\tempurl}


\bibitem[\protect\citeauthoryear{Pratt}{Pratt}{2005}]%
        {ant-quorum}
\bibfield{author}{\bibinfo{person}{Stephen~C. Pratt}.}
  \bibinfo{year}{2005}\natexlab{}.
\newblock \showarticletitle{Quorum sensing by encounter rates in the ant
  Temnothorax albipennis}.
\newblock \bibinfo{journal}{\emph{Behavioral Ecology}} \bibinfo{volume}{16},
  \bibinfo{number}{2} (\bibinfo{year}{2005}), \bibinfo{pages}{488--496}.
\newblock
\urldef\tempurl%
\url{https://doi.org/10.1093/beheco/ari020}
\showDOI{\tempurl}


\bibitem[\protect\citeauthoryear{Pratt, Sumpter, Mallon, and Franks}{Pratt
  et~al\mbox{.}}{2005}]%
        {pratt-paper-1}
\bibfield{author}{\bibinfo{person}{Stephen~C. Pratt},
  \bibinfo{person}{David~J.T. Sumpter}, \bibinfo{person}{Eamonn~B. Mallon},
  {and} \bibinfo{person}{Nigel~R. Franks}.} \bibinfo{year}{2005}\natexlab{}.
\newblock \showarticletitle{An agent-based model of collective nest choice by
  the ant Temnothorax albipennis}.
\newblock \bibinfo{journal}{\emph{Animal Behaviour}} \bibinfo{volume}{70},
  \bibinfo{number}{5} (\bibinfo{year}{2005}), \bibinfo{pages}{1023 -- 1036}.
\newblock
\showISSN{0003-3472}
\urldef\tempurl%
\url{https://doi.org/10.1016/j.anbehav.2005.01.022}
\showDOI{\tempurl}


\bibitem[\protect\citeauthoryear{Reina, Valentini, Fernández-Oto, Dorigo, and
  Trianni}{Reina et~al\mbox{.}}{2015}]%
        {reina-favoring-model}
\bibfield{author}{\bibinfo{person}{Andreagiovanni Reina},
  \bibinfo{person}{Gabriele Valentini}, \bibinfo{person}{Cristian
  Fernández-Oto}, \bibinfo{person}{Marco Dorigo}, {and} \bibinfo{person}{Vito
  Trianni}.} \bibinfo{year}{2015}\natexlab{}.
\newblock \showarticletitle{A Design Pattern for Decentralised Decision
  Making}.
\newblock \bibinfo{journal}{\emph{PLOS ONE}} \bibinfo{volume}{10},
  \bibinfo{number}{10} (\bibinfo{date}{10} \bibinfo{year}{2015}),
  \bibinfo{pages}{1--18}.
\newblock
\urldef\tempurl%
\url{https://doi.org/10.1371/journal.pone.0140950}
\showDOI{\tempurl}


\bibitem[\protect\citeauthoryear{Seeley and Visscher}{Seeley and
  Visscher}{2004}]%
        {bee-quorum}
\bibfield{author}{\bibinfo{person}{Thomas~D. Seeley} {and}
  \bibinfo{person}{P.~Kirk Visscher}.} \bibinfo{year}{2004}\natexlab{}.
\newblock \showarticletitle{Quorum sensing during nest-site selection by
  honeybee swarms}.
\newblock \bibinfo{journal}{\emph{Behavioral Ecology and Sociobiology}}
  \bibinfo{volume}{56}, \bibinfo{number}{6} (\bibinfo{date}{01 Oct}
  \bibinfo{year}{2004}), \bibinfo{pages}{594--601}.
\newblock
\showISSN{1432-0762}
\urldef\tempurl%
\url{https://doi.org/10.1007/s00265-004-0814-5}
\showDOI{\tempurl}


\bibitem[\protect\citeauthoryear{Shahbazi and Barca}{Shahbazi and
  Barca}{2016}]%
        {cockroach-shelter-seeking}
\bibfield{author}{\bibinfo{person}{Hamoon Shahbazi} {and}
  \bibinfo{person}{Jan~Carlo Barca}.} \bibinfo{year}{2016}\natexlab{}.
\newblock \bibinfo{booktitle}{\emph{Cockroach Inspired Shelter Seeking for
  Holonomic Swarms of Flying Robots}}.
\newblock \bibinfo{publisher}{IGI Global}, \bibinfo{address}{Hershey, PA, USA},
  \bibinfo{pages}{687--717}.
\newblock
\showISBNx{9781466695726}


\bibitem[\protect\citeauthoryear{Sims, Humphries, Bradford, and Bruce}{Sims
  et~al\mbox{.}}{2012}]%
        {levy}
\bibfield{author}{\bibinfo{person}{David~W. Sims}, \bibinfo{person}{Nicolas~E.
  Humphries}, \bibinfo{person}{Russell~W. Bradford}, {and}
  \bibinfo{person}{Barry~D. Bruce}.} \bibinfo{year}{2012}\natexlab{}.
\newblock \showarticletitle{Lévy flight and Brownian search patterns of a
  free-ranging predator reflect different prey field characteristics}.
\newblock \bibinfo{journal}{\emph{Journal of Animal Ecology}}
  \bibinfo{volume}{81}, \bibinfo{number}{2} (\bibinfo{year}{2012}),
  \bibinfo{pages}{432--442}.
\newblock
\urldef\tempurl%
\url{https://doi.org/10.1111/j.1365-2656.2011.01914.x}
\showDOI{\tempurl}


\bibitem[\protect\citeauthoryear{Valentini, Ferrante, and Dorigo}{Valentini
  et~al\mbox{.}}{2017}]%
        {n-site-selection-summary}
\bibfield{author}{\bibinfo{person}{Gabriele Valentini}, \bibinfo{person}{Eliseo
  Ferrante}, {and} \bibinfo{person}{Marco Dorigo}.}
  \bibinfo{year}{2017}\natexlab{}.
\newblock \showarticletitle{The Best-of-n Problem in Robot Swarms:
  Formalization, State of the Art, and Novel Perspectives}.
\newblock \bibinfo{journal}{\emph{Frontiers in Robotics and AI}}
  \bibinfo{volume}{4} (\bibinfo{year}{2017}), \bibinfo{pages}{9}.
\newblock
\showISSN{2296-9144}
\urldef\tempurl%
\url{https://doi.org/10.3389/frobt.2017.00009}
\showDOI{\tempurl}


\bibitem[\protect\citeauthoryear{Zhao, Lynch, and Pratt}{Zhao
  et~al\mbox{.}}{2020}]%
        {jiajia-paper}
\bibfield{author}{\bibinfo{person}{Jiajia Zhao}, \bibinfo{person}{Nancy Lynch},
  {and} \bibinfo{person}{Stephen~C. Pratt}.} \bibinfo{year}{2020}\natexlab{}.
\newblock \showarticletitle{A Comprehensive and Predictive Agent-based Model
  for Collective House-Hunting in Ant Colonies}.
\newblock \bibinfo{journal}{\emph{bioRxiv}} (\bibinfo{year}{2020}).
\newblock
\urldef\tempurl%
\url{https://doi.org/10.1101/2020.10.07.328047}
\showDOI{\tempurl}


\end{thebibliography}
